\documentclass[tradiabstract]{aa}
\usepackage{graphicx}
\usepackage{txfonts}
\begin{document}
\title{Search for OB stars running away from young star clusters.\ I.\ NGC\,6611}
\author{V.V.\ Gvaramadze\inst{1,2}
\and D.J.\ Bomans\inst{1}}
\institute{
Astronomical Institute, Ruhr-University Bochum, Universit\"{a}tstrasse 150, 44780
Bochum, Germany\thanks{\email{bomans@astro.rub.de}}
\and
Sternberg Astronomical Institute, Moscow State University,
Universitetskij Pr. 13, Moscow 119992, Russia
\thanks{\email{vgvaram@mx.iki.rssi.ru}}
}
\titlerunning{Bow shocks around young star clusters.\ I.\ NGC\,6611}
\authorrunning{Gvaramadze \& Bomans}
\date{Received 18 June 2008/ Accepted 2 September 2008}
\abstract{N-body simulations have shown that the dynamical decay of
the young ($\sim 1$ Myr) Orion Nebula cluster could be responsible
for the loss of at least half of its initial content of OB stars.
This result suggests that other young stellar systems could also
lose a significant fraction of their massive stars at the very
beginning of their evolution. To confirm this expectation, we used
the Mid-Infrared Galactic Plane Survey (completed by the {\it
Midcourse Space Experiment} satellite) to search for bow shocks
around a number of young ($\la$ several Myr) clusters and OB
associations. We discovered dozens of bow shocks generated by OB
stars running away from these stellar systems, supporting the idea
of significant dynamical loss of OB stars. In this paper, we report
the discovery of three bow shocks produced by O-type stars ejected
from the open cluster NGC\,6611 (M16). One of the bow shocks is
associated with the O9.5Iab star HD165319, which was suggested to be
one of ``the best examples for isolated Galactic high-mass star
formation" (de Wit et al. 2005). Possible implications of our
results for the origin of field OB stars are discussed.}
\maketitle

\section{Introduction}
%
Dynamical three- and four-body encounters between stars in the dense
cores of young star clusters impart to some stars velocities
exceeding the escape velocity from the cluster's potential well
(Poveda et al. \cite{po67}; Leonard \& Duncan \cite{le90}). The
concentration of massive stars towards the center of young clusters
(e.g. Hillenbrand \& Hartmann \cite{hi98}; Gouliermis et al.
\cite{go04}; Chen et al. \cite{ch07}) implies that the dynamical
evolution of cluster cores is dominated by massive stars and that
the majority of escaping stars should be massive. The escaping
massive stars constitute a population of OB field stars (Gies
\cite{gi87}). The escape velocity ranges from several ${\rm km} \,
{\rm s}^{-1}$ (for loose and low-mass clusters) to several tens of
${\rm km} \, {\rm s}^{-1}$ (for compact and massive ones), so that
the typical space velocity of OB field stars is $\sim 10 \, {\rm km}
\, {\rm s}^{-1}$ (e.g. Gies \cite{gi87}). The stars with peculiar
(transverse) velocities exceeding $30 \, {\rm km} \, {\rm s}^{-1}$
are called runaway stars (Blaauw \cite{bl61}). It should be
realized, however, that any star unbound from the parent cluster
should be considered as a `runaway', independently of its peculiar
velocity (cf.  de Wit et al. \cite{de05}). Although the peculiar
velocity of some runaway OB stars could be as high as several $100
\, {\rm km} \, {\rm s}^{-1}$  (e.g. Ramspeck et al. \cite{ra01}) or
even $\sim 500-700 \, {\rm km} \, {\rm s}^{-1}$ (Edelmann et al.
\cite{ed05}; Heber et al. \cite{he08}), the majority have velocities
of several tens of ${\rm km} \, {\rm s}^{-1}$. Therefore, most
runaway OB field stars should be located not far from their birth
clusters.

A (massive) star moving supersonically with respect to the ambient
medium generates a bow shock (Baranov et al. \cite{ba71}; Van Buren
\& McCray \cite{va88}) visible in infrared (Van Buren et al.
\cite{va95}; Noriega-Crespo et al. \cite{no97}; France et al.
\cite{fr07}) and/or ${\rm H}_{\alpha}$ (Kaper et al. \cite{ka97};
Brown \& Bomans \cite{br05}). Observations of runaway OB stars
showed that only a small fraction ($\la 20$ per cent) produce bow
shocks (Van Buren et al. \cite{va95}; Noriega-Crespo et al.
\cite{no97}; Huthoff \& Kaper \cite{hu02}). The main reason for this
is that many runaway stars move through the hot interstellar gas and
their peculiar velocities are lower than the sound speed in the
ambient medium (Huthoff \& Kaper \cite{hu02}). The geometry of a bow
shock allows us to infer the direction of motion of the associated
star and thereby trace its trajectory backwards to the parent
cluster in those cases in which the proper motion of the star has
not been measured directly.

N-body simulations by Pflamm-Altenburg \& Kroupa (\cite{pf06})
showed that the dynamical decay of the young ($\sim 1$ Myr) Orion
Nebula cluster could be responsible for the loss of at least half of
its initial content of OB stars (see also Kroupa \cite{kr04}; cf.
Aarseth \& Hills \cite{aa72}). This result suggests that other young
stellar systems could also lose a significant fraction of their
massive stars at the very beginning of their evolution. We therefore
expect to observe numerous signatures of interaction between ejected
stars and the dense environment of young star clusters. To confirm
this expectation, we searched for bow shocks around a number of
young ($\la$ several Myr) clusters and OB associations using the
Mid-Infrared Galactic Plane Survey. This survey, carried out with
the Spatial Infrared Imaging Telescope onboard the {\it Midcourse
Space Experiment (MSX)} satellite (Price et al. \cite{pr01}), covers
the entire Galactic plane within $|b|<\pm 5\degr$ and provides
images at $18\arcsec$ resolution in four mid-infrared spectral bands
centered at 8.3~$\mu$m (band A), 12.1~$\mu$m (band C), 14.7~$\mu$m
(band D), and 21.3~$\mu$m (band E). To search for possible optical
counterparts to the detected bow shocks, we used the SuperCOSMOS
H-alpha Survey (SHS; Parker et al. \cite{pa05}) and the Digitized
Sky Survey (McLean et al. \cite{mc00}). Our choice of young stellar
systems implies that the majority of ejected stars should not have
traveled far from their birth places (so that the parent clusters
could be identified with more confidence) and that supernova
explosions in massive binaries do not significantly contribute to
the production of runaway stars (Blaauw \cite{bl61}).

We discovered dozens of bow shocks. Three are discussed in this
paper, while the results of study of other objects are presented
elsewhere. The geometry of the three bow shocks suggests that they
are driven by stars expelled from the young open star cluster
\object{NGC\,6611}. In Sect.\,2, we review the relevant data for
this cluster. The results of the search for bow shocks around
NGC\,6611 are presented in Sect.\,3 and discussed in Sect.\,4.
Conclusions are summarized in Sect.\,5.

\section{NGC\,6611}

NGC\,6611 is part of the Ser OB\,1 association in the  Sagittarius
spiral arm. It is embedded in the molecular cloud W37 and is
responsible for ionizing the Eagle Nebula (M16), famous for its
``elephant trunks" and ``EGGs" (evaporating gaseous globules; Hester
et al. \cite{he96}). Numerous distance estimates for NGC\,6611
converge to a figure $d\sim 1.75-2.00$ kpc (Hillenbrand et al.
\cite{hi93}; Loktin \& Beshenov \cite{lo03}; Guarcello et al.
\cite{gu07}; Wolff et al. \cite{wo07}). The age of the cluster is
less well constrained. The presence of four early O-type stars in
the cluster (e.g. Dufton et al. \cite{du06}) implies that its age is
$\la 2-3$ Myr, while the association with the cluster of a B2.5I
star (\object{BD$-13\degr \,4912$}) suggests that star formation in
this part of Ser OB\,1 started $\sim 6$ Myr ago (Hillenbrand et al.
\cite{hi93}; de Winter et al. \cite{de97}; Kharchenko et al.
\cite{kh05}).

\begin{figure*}
\caption{A
$6\degr \times 6\degr$ 21.3~$\mu$m ({\it MSX} band E) image  of
NGC\,6611 and its environments [including two open clusters
\object{NGC\,6604} and \object{NGC\,6618} (M17)], with the positions
of three bow shocks marked by circles. The arrows labeled as L\&B
and Kh show the direction of cluster's peculiar motion, based on the
proper motion measurements, respectively, by Loktin \& Bechenov
(\cite{lo03}) and Kharchenko et al. (\cite{kh05}). The origin of the
arrows corresponds to possible birth places of the cluster $\sim 4$
Myr ago. The thin lines represent the trajectories of ejected stars,
as suggested by the geometry of their bow shocks and proper motion
measurements (see text for details). North is up and east to the
left.}
  \label{NGC}
\end{figure*}

There is a growing body of evidence, however, that clustered star
formation is a rapid process, so the age spread in the cluster
should not exceed $\sim 1$ Myr (e.g. Elmegreen \cite{el00};
Ballesteros-Paredes \& Hartmann \cite{ba07}). The evolutionary
status of the majority of OB stars in NGC\,6611 is consistent with
an age of $\sim 4$ Myr and in the following we adopt this figure to
be the age of the cluster. The observed age spread could be
understood if the most massive (and therefore the youngest) stars in
the cluster are the blue stragglers (the rejuvenated stars formed
via merging of the less massive stars in the course of binary-single
and binary-binary encounters; e.g. Leonard \cite{le95}; Portegies
Zwart et al. \cite{po99}; Gvaramadze \& Bomans \cite{gv08}), while
the older stars are the former field stars trapped in the potential
well of the contracting pre-cluster cloud (Pflamm-Altenburg \&
Kroupa \cite{pf07}). Numerous young ($\sim 0.1-3$ Myr)
pre-main-sequence stars surrounding NGC\,6611 (e.g. Guarcello et al.
\cite{gu07}) could represent the second generation of stars, whose
formation was triggered inside the remainder of the natal molecular
cloud by the activity of massive stars in the cluster.

Radial-velocity measurements for OB stars in NGC\,6611 indicated
that $\simeq 30-50$ per cent are spectroscopic binaries (Bosch et
al. \cite{bo99}). The high percentage of massive binaries is one of
the necessary conditions for production (e.g. via binary-binary
encounters; Mikkola \cite{mi83}; Leonard \& Duncan \cite{le90}) of
runaway OB stars and merged (rejuvenated) massive stars (either
bound or unbound to the cluster; Leonard \cite{le95}). Two other
necessary conditions are the presence in the cluster of a
sufficiently high number of massive binaries (i.e. the cluster
should be massive) and the initially compact structure of the
cluster. It is likely that both conditions were fulfilled in
NGC\,6611, at least at the beginning of its dynamical evolution.

A mass estimate for NGC\,6611 suggests that the cluster could be as
massive as $\simeq 2\times 10^4 \, M_{\odot}$ (Wolff et al.
\cite{wo07}). This estimate was derived by assuming that $\sim 160$
cluster members with masses ranging from 6 to $12 \, M_{\odot}$
constitute $\sim 5$ per cent of the total cluster mass (for a
Salpeter initial mass function). Adopting this mass estimate, we
infer that the cluster should contain $\sim 100$ OB stars, which
exceeds the number of observed stars by a factor of 2.5. It is
possible that the Salpeter initial mass function is not applicable
for NGC\,6611. Another possibility is that the cluster has already
lost most of its massive stars by dynamical ejection (see also
Sect.\,4).

It is likely that NGC\,6611 was initially more centrally
concentrated and that its current size [the radius of the cluster is
$\simeq 7-8$ pc for $d=2$ kpc (Belikov et al. \cite{be00}; Bonato et
al. \cite{bo06})] is the result of dynamical decay of the cluster's
core and overall cluster expansion caused by gas expulsion due to
supernova explosions and stellar winds (e.g. Boily \& Kroupa
\cite{bo03}). The initially compact configuration would be
consistent with the observation that the characteristic radius of
young clusters is $\la 1$ pc, which is independent of their mass
(Kroupa \& Boily \cite{kr02}). For this radius and the above mass
estimate, the escape velocity from the cluster is inferred to have
been $\sim 10 \, {\rm km} \, {\rm s}^{-1}$.

Proper-motion measurements for NGC\,6611 (see Table\,1) suggest that
the cluster is moving in the west-east direction with a peculiar
(tangential) velocity of $\simeq 10-15 \, {\rm km} \, {\rm s}^{-1}$.
To convert the observed proper motion into the true tangential
velocity of the cluster, we used the Galactic constants $R_0 = 8$
kpc and $\Theta _0 =200 \, {\rm km} \, {\rm s}^{-1}$ (e.g. Reid
\cite{re93}; Kalirai et al. \cite{ka04}; Avedisova \cite{av05}), the
solar peculiar motion $(U_{\odot} , V_{\odot} , W_{\odot} )=(10.00,
5.25, 7.17) \, {\rm km} \, {\rm s}^{-1}$ (Dehnen \& Binney
\cite{de98}), and adopted a distance of $d=2$ kpc.

\section{Results of the search}

\subsection{Bow shocks}

We searched for bow shocks around NGC\,6611 in a $12\degr$ wide area
elongated along the Galactic plane and centered on the cluster's
longitude ($l=16\fdg 95$). Along the Galactic latitude, the search
was limited by the {\it MSX} coverage (see Sect.\,1). At a distance
of $1.75-2.00$ kpc, $1\degr$ corresponds to $\simeq 30-34$ pc, so
that potentially we were able to detect bow shocks produced by stars
leaving the cluster at the very beginning of its evolution (i.e.
$\sim 4$ Myr ago) with a peculiar (tangential) velocity of $\la
25-50 \, {\rm km} \, {\rm s}^{-1}$. We realize that some runaway OB
stars were ejected at higher velocities and therefore could be
beyond the area covered by our search. On the other hand, the lower
the peculiar velocity, the lower the percentage of stars moving
supersonically through the ambient medium. Thus, we expect that only
a small fraction of runaway stars located within several degrees
from the cluster are able to generate the bow shocks (see also
Sect.\,4)

The visual inspection of {\it MSX} images revealed several bow-shock
candidates. Three of them (indicated in Fig.\,1 by circles) have a
clear arc-like structure that opens towards NGC\,6611 (Fig.\,2-4),
which suggests that these structures are generated by stars expelled
from the cluster. The bow shocks 1 and 2 are most prominent at
21.3~$\mu$m (Fig.\,2 and Fig.\,3), while the bow shock 3 is visible
only at 8.3~$\mu$m and 21.3~$\mu$m (more clearly at 8.3~$\mu$m;
Fig.\,4).  The bow shocks 2 and 3 have obvious optical counterparts
in the SHS (see Fig.\,3 and Fig.\,4), which are the parts of
$\ion{H} {ii}$ regions, respectively, \object{G\,016.9-01.1} (Kuchar
\& Clark 1997) and \object{RCW\,158} (Rodgers et al. \cite{ro60};
see also Sect.\,4).

\begin{table*}
\caption{Proper-motion measurements for NGC\,6611 and the three
ejected stars. Two measurements are given for each object to
indicate the uncertainties in the measurements. For each data set,
the peculiar (transverse) velocities (in Galactic coordinates) were
calculated and added to the table.} \label{catalog} \centering
\renewcommand{\footnoterule}{}  
\begin{tabular}{lcccccc}
\hline \hline
Object &  $\mu _\alpha \cos \delta$ & $\mu _\delta$ & Sources for & $v_{\rm l}$ & $v_{\rm b}$ \\
~ & mas ${\rm yr}^{-1}$ & mas ${\rm yr}^{-1}$ & proper motions &${\rm km} \, {\rm s}^{-1}$    & ${\rm km} \, {\rm s}^{-1}$  \\
\hline
NGC\,6611 & $0.6\pm 0.1$ & $-0.3\pm 0.1$ & Loktin \& Beshenov (\cite{lo03}) & $9.7\pm0.9$ & $0.7\pm 0.9$ \\
NGC\,6611 & $1.6\pm 0.3$ & $-0.4\pm 0.5$ & Kharchenko et al. (\cite{kh05}) & $12.9\pm 4.3$ & $-7.7\pm 3.5$ \\
star 1 & $0\pm 7$ & $12\pm 3$ & USNO-B1.0 (Monet et al. \cite{mo03}) & $108.7\pm 39.8$ & $61.5\pm 59.7$ \\
star 1 & $-4.3\pm 8.0$ & $-0.9\pm 8.0$ & NOMAD (Zacharias et al. \cite{za04a}) & $-18.2\pm 36.9$ & $39.2\pm 36.9$ \\
BD$-14\degr \, 5040$ & $5.5\pm 1.3$ & $-3.0\pm 1.4$ & NOMAD (Zacharias et al. \cite{za04a}) & $7.5\pm 13.1$ & $-51.7\pm 12.5$ \\
BD$-14\degr \, 5040$ & $7.7\pm 1.6$ & $-4.6\pm 1.8$ & Kharchenko (\cite{kh01}) & $4.0\pm 16.8$ & $-76.5\pm 15.9$\\
HD\,165319 & $-1.2\pm 1.0$ & $-1.4\pm 0.9$ & UCAC2 (Zacharias et al. \cite{za04b}) & $-9.7\pm 8.7$ & $11.0\pm 9.3$\\
HD\,165319 & $0.3\pm 1.0$ & $-1.8\pm 0.6$ & NOMAD (Zacharias et al. \cite{za04a}) & $-6.1\pm 6.7$ & $-3.3\pm 8.7$\\
\hline
\end{tabular}
\end{table*}

\subsection{Associated stars}

The SIMBAD database provides the spectral types of two of the three
stars associated with the bow shocks. We found that  the bow shock 2
is driven by the star BD$-14\degr \, 5040$ [with a photometrically
estimated spectral type of O8V; Kilkenny (\cite{ki93})] and the bow
shock 3 is produced by the star HD\,165319 [whose spectral type of
O9.5Iab was derived spectroscopically by Crampton \& Fisher
(\cite{cr74})]. The bow shock 1 is apparently generated by a star
with coordinates: $\alpha _{2000} =18^{\rm h} 15^{\rm m} 23\fs97,
\delta _{2000} =-13\degr 19\arcmin 35\farcs8$ (hereafter, star\,1;
see Fig.\,2). This star has the visual magnitude of 11.81 (McLean et
al. \cite{mc00}) and 2MASS magnitudes (J,H,${\rm K}_{\rm
s}$)=(8.096,7.684,7.396) (Skrutskie et al. \cite{sk06}).  Using the
extinction law from Rieke \& Lebofsky (\cite{ri85}) and the
photometric calibration of optical and infrared magnitudes for
Galactic O stars by Martins \& Plez (\cite{ma06}), we derived
extinction towards the star\,1 of $A_{\rm V} \simeq 5.6$ mag and
estimated the spectral type of this star to be O9.5III/O5V (for
$d=1.75$ kpc) or O7.5III/O4V ($d=2.00$ kpc). The spectral type of
O7.5III is more consistent with the adopted age of the cluster of 4
Myr. Using the same calibration, we re-estimated the spectral type
of BD$-14\degr \, 5040$ to be 07V ($d=1.75$ kpc) or O6V ($d=2.00$
kpc) and derived extinction of $A_{\rm V} \simeq 3.8$ mag. For both
distances, the isochronal age of the star is less than 4 Myr, so
that this star could be a blue straggler ejected from the cluster
(e.g. Leonard \cite{le95}; Gvaramadze \& Bomans \cite{gv08}). Also,
we estimated the photometric distance to HD\,165319 to be $d\simeq
2.00$ kpc, in good agreement with the distance to NGC\,6611.
\begin{figure}
 \caption{ -- {\it Left:} {\it MSX} 21.3~$\mu$m image of the bow shock 1.
 The position of the associated star (star\,1) is marked by a circle. {\it Right:}
SHS image of the same field with the {\it MSX} 21.3~$\mu$m image overlayed in black
contours. Field size is $8\arcmin$ by $8\farcm5$, the images are
oriented with Galactic ll increasing to the left and Galactic b
increasing upward.}
  \label{bow1}
\end{figure}

\begin{figure}
 \caption{ -- {\it Left:} {\it MSX} 21.3~$\mu$m image of the bow shock 2.
The position of the associated star BD$-14\degr \, 5040$ is marked by a circle.
{\it Right:} SHS image of the same field. Field size and orientation of this figure
are the same as for Fig.\ref{bow1}.}
  \label{bow2}
\end{figure}
\begin{figure}
 \caption{ -- {\it Left:} {\it MSX} 8.3~$\mu$m image of the bow shock 3 and
 the associated star HD\,165319.
{\it Right:} SHS image of the same field. Field size and orientation of
this figure are the same as for Fig.\ref{bow1}.}
  \label{bow3}
\end{figure}

\subsection{Proper motions}

Figures 1, 2 and 4 show that the symmetry axes of the bow shocks 1
and 3 are misaligned with respect to the lines drawn through the
associated stars and the center of the cluster.  This misalignment
could be understood if one takes into account the eastward motion of
the cluster (see Sect.\,2 and Fig.\,1; cf. Hoogerwerf et al.
\cite{ho01}). Assuming that there is no bulk streaming motions of
the ambient gas, we found that the trajectories of the star 1 and
HD\,165319 (suggested by the morphology of their bow shocks)
intersect with the trajectory of the cluster $\simeq 1.6$ Myr ago,
if one uses the cluster proper motion measurements by Loktin \&
Bechenov (\cite{lo03}), or $\sim 0.9$ Myr (star 1) and 1.8 Myr
(HD\,165319) ago for the proper motion measurements by Kharchenko et
al. (\cite{kh05}). In both cases, the stars were ejected more than 2
Myr after the cluster formation. This inference however would be
incorrect if the ambient medium had a significant peculiar velocity,
such that the symmetry axes of the bow shocks did not coincide with
the direction of peculiar motion of the associated stars.

In Table\,1, we provide two representative proper motion
measurements for each star, found with the VizieR Catalogue Service.
One can see that only BD$-14\degr \, 5040$ has a significant proper
motion (i.e. the measurement uncertainties are far less than the
measurements themselves). Both measurements for this star suggest
that it was ejected $\sim 1$ Myr ago with a velocity typical of
classical runaway stars.  It is curious that the symmetry axis of
the bow shock produced by this star points almost exactly towards
the cluster. If however the star was ejected from NGC\,6611 and the
proper motion measurements are correct, then the trajectory of the
star should intersect with the cluster $\simeq 0.30\degr -0.45\degr$
to the west of its present position (for convenience, we show in
Fig.\,1 only one trajectory of the star). The ``incorrect"
orientation of the bow shock, in this case, could be attributed to
the effect of peculiar motion of the ambient gas.

The proper motion measurements for the star\,1 and HD\,165319 are
insignificant. Taken at face value, they suggest that the star\,1 is
a runaway in the classical sense and that it was able to reach its
present position even if ejected from NGC\,6611 only $\sim 0.9$ Myr
ago (see above). The peculiar velocity of HD\,165319 (see Table\,1),
however, is far smaller than the velocity (of $\sim 45-60 \, {\rm
km} \, {\rm s}^{-1}$) inferred from the separation of this star from
the possible trajectories of the cluster and the ejection time
implied by the morphology of its bow shock. This situation cannot be
``improved" if one takes into account that the true (ejection)
velocity of HD\,165319 (i.e. the peculiar velocity in the reference
frame of the cluster) is higher than the ``measured" velocity (given
in the last two columns of Table\,1). Since the available proper
motion measurements for this star are highly unreliable and the
symmetry axis of its bow shock could deviate from the direction of
stellar peculiar motion, one cannot however exclude that either the
true peculiar velocity of HD\,165319 is much higher than the
``measured" one or that this star was ejected soon after the cluster
formation.

\section{Discussion}

Our search for OB stars running away from the young open cluster
NGC\,6611 has resulted in the discovery of three bow shocks
generated by massive stars. The morphology of the bow shocks suggest
that the associated stars were ejected from NGC\,6611. All three
stars are of O-type, which is consistent with the observational fact
that the percentage of O stars among runaways is higher than that of
B stars (e.g. Stone \cite{st91}).

It is natural to assume that the three stars were ejected when the
central density in NGC\,6611 was sufficiently high for close
dynamical encounters between the cluster members to be frequent.
Whether or not the high star density in the cores of young clusters
is primordial (i.e. intrinsic to the cluster formation process; e.g.
Murray \& Lin \cite{mu96}, Clarke \& Bonnell \cite{cl08}) or is the
result of the mass-segregation instability  (e.g. Portegies Zwart et
al. \cite{po99}, G\"urkan et al. \cite{gu04}) remains unclear to
date (e.g. Kroupa \cite{kr08}). In the first case, the cluster could
be collisional at birth, so that one can expect that the majority of
runaway stars were produced at the very beginning of cluster
evolution. In this case, the kinematic age of ejected stars should
be comparable with the age of the cluster. In the  second case, the
beginning of the collisional stage is delayed for several Myr and
the kinematic age of runaway stars is younger than the age of the
parent cluster. Accurate proper motion measurements of runaway stars
and their parent clusters are needed to distinguish between these
two possibilities.

The large separation between HD\,165319 and NGC\,6611 and the low
peculiar velocity of the star might imply that it was ejected soon
after the cluster formation, supporting the first possibility. The
proper motion measurements for this star are however unreliable (see
Table\,1), which leaves a possibility that HD\,165319 was ejected
recently, as suggested by the symmetry axis of its bow shock. The
late production of runaway stars also follows from the young
kinematic age of BD$-14\degr \, 5040$ (the only one of the three
stars with reliable proper motion measurements) and might be
inferred from the morphology of the bow shock produced by star\,1.
The available astrometric data, however, do not allow us to
differentiate between whether or not NGC\,6611 experienced a burst
of star ejection $\sim 1-2$ Myr ago or this process continued up
until cluster formation. Indirect support of the latter possibility
comes from the study of field O stars by Schilbach \& R\"{o}ser
(\cite{sc08}). These authors retraced the orbit of the O-type star
\object{HD\,157857} in the Galactic potential and found that this
star was ejected from NGC\,6611 $\simeq 3.8$ Myr ago with a peculiar
velocity of $\simeq 114 \, {\rm km} \, {\rm s}^{-1}$. More robust
proper motion measurements for star 1 and HD\,165319 would help to
solve the problem.

By assuming that all three stars were ejected from NGC\,6611 and
taking into account that $\la 20$ per cent of runaway stars produce
bow shocks (see Sect.\,1), one can infer that at least 15 massive
stars have been lost by the cluster during its lifetime. The true
number of ejected OB stars could be far higher since high-velocity
runaways would have already traveled beyond the region covered by
the {\it MSX} survey\footnote{A possible example of such a
high-velocity runaway is the aforementioned star HD\,157857, located
at $\sim 13\degr$ from NGC\,6611.}, while the more numerous stars
ejected at low velocities ($\la 30 \, {\rm km} \, {\rm s}^{-1}$)
have lower probabilities of their motion relative to the ambient
medium being supersonic and therefore of being able to generate the
bow shocks. We therefore believe that the percentage of stars
producing bow shocks could be far lower than 20 per cent, based on
the study of ``classical" runaway stars. From this follows that the
number of ejected OB stars could be comparable with or higher than
the number still residing in NGC\,6611, which could explain the
apparent deficit of OB stars in the cluster (see Sect.\,2).

In addition to the dynamical ejection, there are two other effects
that can reduce the size of the OB star population in the cluster:
(i) supernova explosions of the most massive stars and (ii) merging
of stars in the  course of close binary-single and binary-binary
star encounters. Supernova explosions in massive binaries could be
accompanied by ejection of companion stars (e.g. Blaauw \cite{bl61};
Leonard \& Dewey \cite{le93}), which in addition reduce the
population of massive stars in the cluster. Supernova explosions
begin to occur in clusters of age $\ga 3$ Myr. Flagey (\cite{fl07})
revealed a shell-like structure within the Eagle Nebula and
suggested that this shell was a young ($\sim 3000$ yr) supernova
remnant. If confirmed, this suggestion would imply that NGC\,6611 is
at least as old as 3 Myr, which is consistent with the age of the
cluster adopted in Sect.\,2. For the age and mass of NGC\,6611 of
$\simeq 4$ Myr and $\simeq 2\times 10^4 \, M_{\odot}$, respectively,
and assuming a Salpeter initial mass function, one finds that
$\simeq 10$ stars have exploded as supernovae since the cluster
formation. Correspondingly, the number of massive runaways produced
by binary disruptions following supernova explosions cannot exceed
this number. Merging of stars is a by-product of star ejection by
means of binary-single and binary-binary encounters (e.g. Leonard
1995). Four early-type O stars in NGC\,6611 and the star producing
the bow shock\,2 (BD$-14\degr \, 5040$) appear younger than the main
population of OB stars in the cluster and therefore could be the
rejuvenated merged stars (e.g. Leonard \cite{le95}; Portegies Zwart
et al. \cite{po99}; Gvaramadze \& Bomans \cite{gv08}). Therefore, we
suggest that at least 5 OB stars were ``lost" by the cluster via
merging.

We note that HD\,165319 was considered by de Wit et al.
(\cite{de05}) to be one of ``the best examples for isolated Galactic
high-mass star formation". de Wit et al. (\cite{de04}, \cite{de05})
searched for parent star clusters for 43 O-type field stars to
confirm whether or not these massive stars originated in a clustered
mode of star formation. They found that 11 stars from their sample
cannot  be associated with a nearby cluster or star-forming region
and suggested that $\sim 4\pm2$ per cent of all field O-type stars
could be formed in isolation (cf. Chu \& Gruendl \cite{ch08}). If
confirmed, this result could have important implications for
understanding massive star formation. It would either imply the
existence of a special mode of isolated star formation (e.g. Yorke
\& Sonnhalter \cite{yo02}; Li et al. \cite{li03}) or allow the
possibility that low mass ($< 100 \, M_{\odot}$) clusters are able
to form O stars (e.g. Elmegreen \cite{el06}; Parker \& Goodwin
\cite{pa07}).

de Wit et al. (\cite{de05}) restricted the search for parent
clusters within a circle of radius 65 pc; this figure comes from the
average peculiar radial velocity of field O stars, which is $6.4 \,
{\rm km} \,{\rm s}^{-1}$ (Gies \cite{gi87}), multiplied by their
average lifetime of 10~Myr ($1 \ \, {\rm km} \, {\rm s}^{-1} \,
\times$ 1~Myr $\simeq 1$ pc). It is possible therefore that at least
some of the 11 stars, suggested by de Wit et al. (\cite{de05}) to be
formed in isolation, were instead ejected from clusters more distant
than 65 pc. The discovery of the bow shock around HD\,165319 and
identification of its possible parent cluster (separated by $\simeq
105$ pc) lends support to this suggestion. We note that some of the
remaining 10 stars could be rejuvenated stars (blue stragglers)
formed via merging of less massive stars in binaries ejected from
clusters located far beyond the circle of radius of 65 pc. Study of
field O-type stars by Schilbach \& R\"{o}ser (\cite{sc08}) suggests
that 6 of these 10 stars were ejected from open star clusters and
were therefore formed in the clustered way. In some cases, however,
the times-of-flight derived by Schilbach \& R\"{o}ser (\cite{sc08})
exceed the lifetimes of the stars. The proposed identification of
the parent clusters would therefore be incorrect if these stars are
not blue stragglers.

We searched for bow shocks around the remaining 10 O-type stars
formed in `isolation' using the {\it MSX} survey but none were found
(note that the survey covers only 8 of 10 stars). This non-detection
could be due to the limited and inhomogeneous sensitivity of the
survey (see Price et al. \cite{pr01}). It is also possible that some
of these stars have not produced bow shocks simply because their
peculiar velocities are less than the sound speed in the (hot)
ambient interstellar medium (see Sect.\,1). We note that the
peculiar velocities relative to the ambient medium (and therefore
ability to generate bow shocks) of some runaway stars could be
affected by the cluster peculiar motion. A (massive) star ejected in
the direction opposite to the cluster's motion will have a reduced
peculiar velocity relative to the local standard of rest, while its
true (ejection) velocity with respect to the cluster could be
sufficiently high to classify the star as a ``classical" runaway.
This star has not only a lower probability of producing a bow shock,
but is also more distant from the parent cluster than one can infer
from its apparent proper motion. Since most ejected stars have
velocities of several tens ${\rm km} \, {\rm s}^{-1}$ (i.e.
comparable with the escape velocity from the parent cluster) and the
peculiar velocities of clusters could be as high as $\sim 10-20 \,
{\rm km} \, {\rm s}^{-1}$ (de Zeeuw et al. \cite{de99}; see also
Sect.\,2), the above effect could be significant and might cause an
excess of bow-shock-producing stars ahead of the moving cluster.

\begin{figure} 
\caption{A $0\fdg5 \times 0\fdg5$ SHS image of RCW\,158,
again oriented in Galactic coordinates
(increasing ll to left, and increasing b upward).}
  \label{RCW}
\end{figure}
In Sect.\,3.1, we mentioned that the bow shock 3 is part of the
$\ion{H} {ii}$ region RCW158. de Wit et al. (\cite{de05}) consider
this nebula as the tracer of isolated high mass star formation. The
almost central location of HD\,165319 in RCW\,158 (Fig.\,5)
suggests, however, that RCW\,158 might be a Str\"{o}mgren zone
created by the ultraviolet emission of the star. We now check this
possibility. Assuming that the gas in RCW\,158 is fully ionized and
its temperature is $10^4$ K, one can obtain an estimate of the gas
number density (e.g. Lequeux \cite{le05})
\begin{equation}
n\simeq 22 \, {\rm cm}^{-3} \, \left({R_{\rm S} \over 6.88 \, {\rm pc}}
\right)^{-3/2} \left({S(0) \over 4.90\times 10^{48} \, {\rm photons} \,
{\rm s}^{-1}} \right)^{1/2} \, ,
\end{equation}
where $R_{\rm S}$ is the Str\"{o}mgren radius (derived from the
angular radius of RCW\,158 of $\sim 12\arcmin$ and $d=2$ kpc) and
$S(0)$ is the total ionizing-photon luminosity of an O9.5Iab star
(taken from Martins et al. \cite{ma05}). This value should be equal
to that derived from the ram pressure balance between the stellar
wind and the ambient medium:
\begin{eqnarray}
n \simeq 17 \, {\rm cm}^{-3} \, \left({\dot{M} \over 5.3\times 10^{-6} \,
M_{\odot} {\rm yr}^{-1}}\right) \left({v_{\infty} \over 2100 \, {\rm km}
\, {\rm s}^{-1}}\right)  \ \nonumber \\
\times \left({R_0 \over 0.8 {\rm pc}}\right)^{-2}
\left({v_{\ast} \over 50\, {\rm km} \, {\rm s}^{-1}}\right)^{-2} \, ,
\end{eqnarray}
where $\dot{M}$ and $v_\infty$ are the wind mass-loss rate and
terminal velocity of an O9.5Iab star (taken from Mokiem et al.
\cite{mo07}), $R_0$ is the stand-off radius\footnote{We neglected
here a geometric factor of order unity, taking into account the
inclination of the bow shock due to the peculiar radial velocity of
the star of $\simeq 20.5 \, {\rm km} \, {\rm s}^{-1}$ [derived from
the measured radial velocity of $\simeq 25.4 \, {\rm km} \, {\rm
s}^{-1}$ (Kharchenko et al. \cite{kh07})].}, and $v_{\ast}$ is the
velocity of the star relative to the ambient medium, assumed to be
equal to $50\, {\rm km} \, {\rm s}^{-1}$ (see Sect.\,3.3). The good
agreement between the two estimates supports our suggestion that
RCW\,158 is the Str\"{o}mgren zone of HD\,165319. The stellar
parameters used in Eq.\,2 could however still vary within the
classification [see e.g. the values given by Markova et al.
(\cite{ma04}) for the same spectral and luminosity class]. Using
different possible values for the stellar parameters and allowing
the velocity to be between 30 km s$^{-1}$ and 60 km s$^{-1}$ in
Eq.\,2, we estimate the approximate lower and upper bounds of the
density to be 10 cm$^{-3}$ and 30 cm$^{-3}$, respectively. This
density range is again in good agreement with the density estimate
from the Str\"omgren formula (Eq.\,1). Further support comes from
the observation that RCW\,158 is partially surrounded by 8.3~$\mu$m
filamentary structures, which are typical of photodissociation
regions (Hollenbach \& Tielens \cite{ho97}). We therefore conclude
that the ionizing photons emitted by HD\,165319 have created a
Str\"{o}mgren sphere in the dense ambient medium (visible as
RCW\,158), while the interaction between the wind of this
supersonically moving star and the ambient gas generates a bow shock
ahead of the star and evacuates a cavity behind (see Fig.\,5; cf.
Raga et al. \cite{ra97}).

\section{Conclusions}

With the detection of three bow-shock-producing stars in the
vicinity of NGC\,6611, we have found strong evidence that the
proposed loss of massive stars, due to dynamical processes in the
early evolution of young clusters, is indeed operating. While
existing astrometric data for the three stars do not allow us to
determine precisely the timing of the ejection or study the
interaction of the runaway stars with the ISM in detail, the
morphology of the bow shocks and the proper motion of the cluster
itself allows us to argue that these stars originated in NGC\,6611.

Clearly, our finding opens up a number of possibilities for
follow-up investigations of the NGC\,6611 bow shocks and the ejected
stars. More accurate proper motion measurements will allow us to
check if the stars were ejected continuously or during a short
period of cluster evolution. The most important point nevertheless
is that the idea proposed by Pflamm-Altenburg \& Kroupa
(\cite{pf06}) is supported by direct observation, at least in the
case of NGC\,6611. The results for several other young clusters will
be presented in forthcoming papers.

\begin{acknowledgements}

The authors are grateful to P.\ Kroupa, F.\ Martins and J.\
Pflamm-Altenburg for interesting discussions, to the anonymous
referee for suggestions allowing us to improve the manuscript and to
N.\ Flagey for sending us his PhD Thesis. This research has made use
of the NASA/IPAC Infrared Science Archive, which is operated by the
Jet Propulsion Laboratory, California Institute of Technology, under
contract with the National Aeronautics and Space Administration, the
SIMBAD database and the VizieR catalogue access tool, both operated
at CDS, Strasbourg, France. The authors acknowledge financial
support from the Deutsche Forschungsgemeinschaft (grants 436 RUS
17/104/06 and BO 1642/14-1) for research visits of VVG at the
Astronomical Institute of the Ruhr-University Bochum.
\end{acknowledgements}


\begin{thebibliography}{}
%
\bibitem[1972]{aa72} Aarseth, S.J., \& Hills, J.G. 1972, A\&A, 21, 255
\bibitem[2005]{av05} Avedisova, V.S. 2005, Astronomy Reports, 49, 435
\bibitem[2007]{ba07} Ballesteros-Paredes, J., \& Hartmann, L. 2007, \rmxaa, 43, 123
\bibitem[1971]{ba71} Baranov, V.B., Krasnobaev, K.V., \& Kulikovskii, A.G. 1971, Soviet Phys. Dokl., 15, 791
\bibitem[2000]{be00} Belikov, A.N., Kharchenko, N.V., Piskunov, A.E., \& Schilbach, E. 2000, A\&A, 358, 886
\bibitem[1961]{bl61} Blaauw, A. 1961, Bull. Astron. Inst. Netherlands, 15, 265
\bibitem[2003]{bo03} Boily, C.M., \& Kroupa, P. 2003, MNRAS, 338, 665
\bibitem[2006]{bo06} Bonatto, C., Santos, Jr., J.F.C., \& Bica, E. 2006, A\&A, 445, 567
\bibitem[1999]{bo99} Bosch, G.L., Morrell, N.I., \& Niemel\"{a}, V.S. 1999, Rev. Mex. Astron. Astrofis., 35, 85
\bibitem[2005]{br05} Brown, D., \& Bomans, D.J. 2005, A\&A, 439, 183
\bibitem[2007]{ch07} Chen, L., de Grijs, R., \& Zhao, J.L. 2007, AJ, 134, 1368
\bibitem[2008]{ch08} Chu, Y.-H., \& Gruendl, R.A. 2008, in Massive Star Formation: Observations Confront Theory, eds. H. Beuther, H. Linz, \& T. Henning,
ASP Conf. Ser., 387, 415
\bibitem[2008]{cl08} Clarke, C.J., \& Bonnell, I.A. 2008, MNRAS, 388, 1171
\bibitem[1974]{cr74} Crampton, D., \& Fisher, W.A. 1974, Publ. Dom. Astrophys. Obs., 14, 283
\bibitem[1997]{de97} de Winter, D., Koulis, C., Th\'{e}, P.S., et al. 1997, \aaps , 121, 223
\bibitem[2004]{de04} de Wit, W.J., Testi, L., Palla, F., Vanzi, L., \& Zinnecker, H. 2004, A\&A, 425, 937
\bibitem[2005]{de05} de Wit, W.J., Testi, L., Palla, F., \& Zinnecker, H. 2005, A\&A, 437, 247
\bibitem[1999]{de99} de Zeeuw, P.T., Hoogerwerf, R., de Bruijne, J.H.J., Brown, A.G.A., \& Blaauw, A. 1999, AJ, 117, 354
\bibitem[1998]{de98} Dehnen, W., \& Binney, J.J. 1998, MNRAS, 298, 387
\bibitem[2006]{du06} Dufton, P.L., Smartt, S.J., Lee, J.K., et al. 2006, A\&A, 457, 265
\bibitem[2005]{ed05} Edelmann, H., Napiwotzki, R., Heber, U., Christlieb, N., \& Reimers, D. 2005, ApJ, 634, L181
\bibitem[2000]{el00} Elmegreen, B.G. 2000, ApJ, 530, 277
\bibitem[2006]{el06} Elmegreen, B.G. 2006, ApJ, 648, 572
\bibitem[2007]{fl07} Flagey, N. 2007, PhD Thesis, Universit\'{e} Paris Sud, 258 p.
\bibitem[2007]{fr07} France, K., McCandliss, S.R., \& Lupu, R.E. 2007, ApJ, 655, 920
\bibitem[1987]{gi87} Gies, D.R. 1987, ApJS, 64, 545
\bibitem[2004]{go04} Gouliermis, D., Keller, S.C., Kontizas, M., Kontizas, E., \& Bellas-Velidis, I. 2004, A\&A, 416, 137
\bibitem[2007]{gu07} Guarcello, M.G., Prisinzano, L., Micela, G., et al. 2007, A\&A, 462, 245
\bibitem[2004]{gu04} G\"{u}rkan, M.A., Freitag, M., \& Rasio, F.A. 2004, ApJ, 604, 632
\bibitem[2008]{gv08} Gvaramadze, V.V., \& Bomans, D.J. 2008, A\&A, 485, L29
\bibitem[2008]{he08} Heber, U., Edelmann, H., Napiwotzki, R., Altmann, M., \& Scholz, R.-D. 2008, A\&A, 483, L21
\bibitem[1996]{he96} Hester, J.J., Gilmozzi, R., O'Dell, C.R., et al. 1996, ApJ, 369, L75
\bibitem[1998]{hi98} Hillenbrand, L.A., \& Hartmann, L.W. 1998, ApJ, 492, 540
\bibitem[1993]{hi93} Hillenbrand, L., Massey, P., Strom, S.E., \& Merrill, K.M. 1993, AL, 106, 1906
\bibitem[1997]{ho97} Hollenbach, D.J., \& Tielens, A.G.G.M. 1997,ARA\&A, 35, 179
\bibitem[2001]{ho01} Hoogerwerf, R., de Bruijne, J.H.J., \& Zeeuw, P.T. 2001, A\&A, 365, 49
\bibitem[2002]{hu02} Huthoff, F., \& Kaper, L. 2002, A\&A, 383, 999
\bibitem[2004]{ka04} Kalirai, J.S., Richer, H.B., Hansen, B.M., et al. 2004, 601, 277
\bibitem[1997]{ka97} Kaper, L., van Loon, J.Th., Augusteijn, T.,  et al. 1997, ApJ, 475, L37
\bibitem[2001]{kh01} Kharchenko, N.V. 2001, Kinematika Fiz. Nebesnykh Tel, 17, 409
\bibitem[2005]{kh05} Kharchenko, N.V., Piskunov, A.E., R\"{o}ser, S., Schilbach, E., \& Scholz, R.-D. 2005, A\&A, 438, 1163
\bibitem[2007]{kh07} Kharchenko, N.V., Scholz, R.-D., Piskunov, A.E., R\"{o}ser, S., \& Schilbach, E. 2007, AN, 328, 889
\bibitem[1993]{ki93} Kilkenny, D. 1993, S. Afr. Astron. Obs. Circ., 15, 53
\bibitem[2004]{kr04} Kroupa, P. 2004, New Astron. Rev., 48, 47
\bibitem[2008]{kr08} Kroupa, P. 2008, preprint (astro-ph/0803.1833)
\bibitem[2002]{kr02} Kroupa, P., \& Boily, C.M. 2002, MNRAS, 336, 1188
\bibitem[1997]{ku97} Kuchar, T.A., \& Clark, F.O. 1997, ApJ, 488, 224
\bibitem[2005]{le05} Lequeux, J. 2005, The Interstellar Medium (Berlin: Springer)
\bibitem[1991]{le91} Leonard, P.J.T. 1991, AJ, 101, 562
\bibitem[1995]{le95} Leonard, P.J.T. 1995, MNRAS, 277, 1080
\bibitem[1993]{le93} Leonard, P.J.T., \& Dewey, R.J. 1993, in Luminous High-Latitude Stars, ed. D. Sasselov, ASP Conf. Ser., 45, 239
\bibitem[1990]{le90} Leonard, P.J.T., \& Duncan, M.J. 1990, AJ, 99, 608
\bibitem[2003]{li03} Li, Y., Klessen, R.S., \& Mac Low, M.-M. 2003, ApJ, 592, 975
\bibitem[2003]{lo03} Loktin, A.V., \& Beshenov, G.V. 2003, Astron. Rep., 47, 6
\bibitem[2004]{ma04} Markova, N., Puls, J., Repolust, T., \& Markov, H. 2004, A\&A, 413, 693
\bibitem[2006]{ma06} Martins, F., \& Plez, B. 2006, A\&A, 457, 637
\bibitem[2005]{ma05} Martins, F., Schaerer, D., \& Hillier, D. J. 2005, A\&A, 436, 1049
\bibitem[2000]{mc00} McLean, B.J., Greene, G.R., Lattanzi, M.G., \& Pirenne, B. 2000, in Astronomical Data Analysis Software and Systems IX, eds. N. Manset, C. Veillet, \& D. Crabtree, ASP Conf. Ser., 216, 145
\bibitem[1983]{mi83} Mikkola, S. 1983, MNRAS, 205, 733
\bibitem[2007]{mo07} Mokiem, M.R., de Koter, A., Vink, J.S., et al. 2007, A\&A, 473, 603
\bibitem[2003]{mo03} Monet D.G., Levine S.E., Casian B., et al. 2003, AJ, 125, 984
\bibitem[1996]{mu96} Murray, S.D., \& Lin, D.N.C. 1996, ApJ, 467, 728
\bibitem[1997]{no97} Noriega-Crespo, A., Van Buren, D., \& Dgani, R. 1997, AJ, 113, 780
\bibitem[2007]{pa07} Parker, R.J., \& Goodwin, S.P. 2007, MNRAS, 380, 1271
\bibitem[2005]{pa05} Parker, Q.A., Phillipps, S., Pierce, M.J., et al. 2005, MNRAS, 362, 689
\bibitem[1999]{po99} Portegies Zwart, S.F., Makino, J., McMillan, S.L.W., \& Hut, P. 1999, A\&A, 348, 117
\bibitem[1967]{po67} Poveda, A., Ruiz, J., \& Allen, C. 1967, Bol. Obs. Tonantzintla Tacubaya, 4, 86
\bibitem[2006]{pf06} Pflamm-Altenburg, J., \& Kroupa, P. 2006, MNRAS, 373, 295
\bibitem[2007]{pf07} Pflamm-Altenburg, J., \& Kroupa, P. 2007, MNRAS, 375, 855
\bibitem[2001]{pr01} Price, S.D., Egan, M.P., Carey, S.J., Mizuno, D.R., Kuchar, T.A. 2001, AJ, 121, 2819
\bibitem[1997]{ra97} Raga, A.C., Noriega-Crespo, A., Cant\'{o}, J., et al. 1997, \rmxaa , 33, 73
\bibitem[2001]{ra01} Ramspeck, M., Heber, U., \& Moehler, S. 2001, A\&A, 378, 907
\bibitem[1993]{re93} Reid, M.J. 1993, ARA\&A, 31, 345
\bibitem[1985]{ri85} Rieke, G.H., \& Lebofsky, M.J. 1985, ApJ, 288, 618
\bibitem[1960]{ro60} Rodgers, A.W., Campbell, C.T., \& Whiteoak, J.B. 1960, MNRAS, 121, 103
\bibitem[2008]{sc08} Schilbach, E., \& R\"{o}ser, S. 2008, A\&A, in press (astro-ph/0806.0762)
\bibitem[2006]{sk06} Skrutskie, M.F., Cutri, R.M., Stiening, R. et al. 2006, AJ, 131, 1163
\bibitem[1991]{st91} Stone, R.C. 1991, AJ, 102, 333
\bibitem[1988]{va88} Van Buren, D., \& McCray, R. 1988, ApJ, 329, L93
\bibitem[1995]{va95} Van Buren, D., Noriega-Crespo, A., \& Dgani, R. 1995, AJ, 110, 2914
\bibitem[2007]{wo07} Wolff, S.C., Strom, S.E., Dror, D., \& Venn, K. 2007, AJ, 133, 1092
\bibitem[2002]{yo02} Yorke, H.W., \& Sonnhalter, C. 2002, ApJ, 569, 846
\bibitem[2004a]{za04a} Zacharias, N., Monet, D. G., Levine, S. E., et al. 2004a, AAS, 205, 4815
\bibitem[2004b]{za04b} Zacharias, N., Urban, S.E., Zacharias, M.I., et al. 2004b, AJ, 127, 3043

\end{thebibliography}
\end{document}